# Development and properties of composite cement reinforced coconut fiber with the addition of fly ash

N. I. M. Nadzri [1,2*], J. B. Shamsul,[1,2], M. N. Mazlee[1,2]

[1]*Sustainable Engineering Research Cluster, Universiti Malaysia Perlis*
*Blok B, Taman Pertiwi Indah, Seriab*
*01000 Kangar, Perlis, Malaysia*
[2]*School of Materials Engineering, Universiti Malaysia Perlis*
*02600 Arau, Perlis, Malaysia*

[*]izzati.nadzri@yahoo.com, sbaharin@unimap.edu.my, mazlee@unimap.edu.my

**ABSTRACT**

In this paper, the effectiveness utilization of agriculture wastes and industrial wastes in the composites cement has been studied in terms of physical and mechanical properties. Twenty weight percent of fly ash and eighty weight percent of sand were added to the composite cement. The different weight percent of coconut fiber (3 wt. %, 6 wt. %, 9 wt. %, 12 wt. % and 15 wt. %) was added in the composition as reinforcement for cement composites. Water to cement ratio ranging from 0.55 to 0.70 was added into the cement composites accordingly to maintain its workability. Then the cement composites were cured in water for 7, 14 and 28 days respectively. Results for physical properties (density, moisture content, water absorption) and mechanical properties (compressive strength, modulus of rupture) are presented.

**INTRODUCTION**

Cement reinforced natural fibers composites with higher fiber content are developed for equivalent performance of glass fiber reinforcement. Its characteristics in term of density, cost and preserving the environment make natural fiber in cement composites to be a potential replacement of glass fibers in many applications that do not require very high load bearing capabilities [1,2]. They also exhibit better toughness, enhanced cracking behavior and improved fatigue that have been documented by several researchers [3,4]. Coconut fiber is particularly selected as reinforcement in the cement composite. Coconut fibers are also known as coir fiber is abundantly found in Brazil, China, Malaysia and Thailand [5]. Studies on the comparison between four fibers (coconut fiber, jute, sisal and hibiscus) have been made by subjected these four fibers to impact loading. It showed that the addition of natural fiber has increased 3 to 18 times of impact resistance of cement composite. Coconut fibers have one of the lowest water absorption value compared to sisal and jute fibers. This is because of cement reinforced with coconut fiber composites have absorbed the highest impact energy and proved to have a better performance [6]. The urge of preserving the natural resources such as sand has becoming the main issue in the cement composite industry. A reasonable solution is to use fly ash and natural fiber in the cement composite industry without sacrificing its properties. Vast researches have been conducted on application of natural fibers or fly ash in development of construction materials and limited studies have been made in combining these two wastes in one system [7,8]. In this present study, the investigation has focused on the physical and mechanical properties of the cement reinforced coconut fiber composite with the addition of fly ash. This research work was aimed to investigate the potential use of fly ash in substituting the portion of sand and coconut fiber as reinforcement in the production of cement composite.

**EXPERIMENTAL PROCEDURES**

**Materials and Samples Preparation:**

The mix proportions of the composite are given in Table 1. The cement used was Ordinary Portland Cement (Blue Lion trademark), fly ash type F, sand, coconut fiber and water were used in this work. The ratio of cement, sand and fly ash were kept constant at 1 : 0.2 : 0.8. The water to cement ratio was varied according to the percentage of coconut fiber added in the mixture. Coconut fiber was weighed according to the percentage ratio of cement weight.

**Table 1**: Mix proportions of cement composite.

| Sample | Water/ Cement | Coconut fiber (wt. %) |
|---|---|---|
| 1 | 0.55 | 0 |
| 2 | 0.55 | 3 |
| 3 | 0.60 | 6 |
| 4 | 0.60 | 9 |
| 5 | 0.65 | 12 |
| 6 | 0.70 | 15 |

The work has focused on six different percentage of coconut fiber to cement weight. Then fly ash was added in the mixture as replacement for sand. The weight percent starts from the reference sample which was 0 wt. % of coconut fiber and then increased to 3 wt. %, 6 wt. %, 9 wt. %, 12 wt. % and 15 wt. %. All the raw materials were added in a mechanical mixer until slurry mixture was formed. Enough amount of water and the speed of the mixer are crucial during mixing to provide better workability of fly ash and to avoid agglomeration of coconut fiber. The uniform slurry formed was then placed in the mould according to the mould's size. The size of the mould is depending on the type of test that was carried out for the composite samples. The samples were kept in the mould for 24 hours before it was demoulded and cured for 7, 14, and 28 days of curing.

**Testing Procedures**

The samples were taken care to avoid any contaminations and damage that would affect the result. Density test was carried out by measuring its mass and volume for different curing days. Samples for water absorption test were immersed in a water bath for 24 hours and the weight difference of before and after immersion was recorded. For moisture content test, samples were placed in an oven at 100°C for 24 hours and the difference of samples weight before and after samples were placed in an oven was measured. These three test's samples used the same size of mould, which is 100 mm x 100 mm x 40 mm. Compressive strength of the composites under crushing load was determined by using the Instron machine with cross-head speed at 3 mm/min. compression test. The sample size used was 160 mm x 40 mm x 40 mm. The size of the sample was 400 mm x 100 mm x 16 mm. These five tests have complied with BS 5669 Part 1: 1989.

**RESULTS AND DISCUSSION**

**Density**

Density of the composites is summarized in Figure 1a. The highest density value of 2245 kg/m$^3$ was exhibited by the cement composite without the presence of fly ash after 7 days of curing. Cement composite with 15 wt. % coconut fiber after 7 days show the lowest value (1735 kg/m$^3$) of density. In term of coconut fiber content, the value of density decreases with the increasing content of coconut fiber. As reported by Aggrawal [9], the replacement of cement or sand (dense materials) by natural fiber (lighter material) resulted in an increase in total volume of the mixture. Further, the increase in volume of the mixture is directly related to the increase in natural fiber content. This increase in volume of the mixture resulted in a decrease in density of the samples. In general, the result obtained is similar to the previous works, which reported that density values of cement composite decreased with the increasing content of natural fiber [10, 11].

**Moisture Content**

Figure 1b depicts the comparison of moisture content of the composites after 7, 14, and 28 days of curing. From the graph, it was indicated that the composite with 6 wt. % of coconut fiber after 28 days of curing gives highest moisture content, which is 9.31 % and the lowest moisture content of 3.64 % is given by the reference sample after 7 days. Percent of moisture content increased from 0 wt. % to 6 wt. % of coconut fiber, but sharply decreased at 9 wt. % of coconut fiber. Finally, the percent of moisture content gradually increases as it reaches to 15 wt. % coconut fiber. As the number of curing days increased, the value of moisture content also shows an increment.

Generally, the moisture content results indicate similar trend of increasing moisture content when cement reinforced with more than 9 wt. % of coconut fiber. However, when the addition of coconut fiber is less that 9 wt. %, it shows no significant change in the moisture content. Fly ash addition as sand replacement in the composites also contributes a factor in the decreasing value of moisture content. Higher moisture content will lead to lower in mechanical properties generally, and compressive strength in particular [10,12].

**Water Absorption**

Water absorption results for composites after 7, 14, and 28 days of curing are presented in Figure 1c. Water absorption increased with increases weight percent of coconut fiber. The highest value of water absorption is exhibited by the composite cement reinforced with 15 wt. % coconut fiber (5.67 %) after 14 days, while the reference sample for 28 days gives the lowest value (0.79 %) of water absorption in composite cement. The value of water absorption gradually increases from 7 days to 14 days, but experienced a sudden drop when the curing day takes place after 28 days. These apply to all composite cement reinforced coconut fiber.

The same pattern of water absorption is observed in the previous research conducted by Alida et al. [10] but higher in value of water absorption. Fly ash, that develop a pozzolanic properties also have similar characteristics of coconut fiber, which require high amount of water for workability. Composite cement produced using more coconut fiber content exhibited the highest water absorption that attributed to their low density and hence higher porosity in the composite cement [13]. Asasutjarit et al. [14] reported that low density of composite cement reinforced with natural fibers have more void spaces than dense ones so that more water can be absorbed.

**Compressive Strength**

The behavior of composite cement under crushing load is determined by a compression test. The compressive strength of the composite cement for all composites is depicted in Figure 1d. It was indicated that the composite cement reinforced with 9 wt. % of coconut fiber after 14 days has the highest compressive strength which is 58.98 MPa and the lowest compressive strength that is 21.19 MPa which is obtained from the composite cement with 12 wt. % of coconut fiber after 28 days of curing.

The composite cement with higher content of fiber suffered some loss of strength at later ages, that attributed to the possibility that the water absorbed could not utilized for further hydration thus leading to voids in the composite cement [15]. The effect of fly ash on the compressive strength can be seen when this industrial waste replaces proportion of sand. Basar and Aksoy [16] reported that increased proportions of replaced sand with waste foundry sand (fine aggregate) resulted in a reduction in compressive strength due to the higher surface area of fine particles, which led to the reduction of water cement gel in matrix. Therefore, binding of aggregates with cement will be difficult to take place. Fly ash causes a loose of contact between the cement and sand. Nevertheless, the composites cement exhibited similar results to the reference sample.

**Modulus of Rupture (MOR)**

The MOR results are shown in Figure 1e, where the maximum MOR for all the composites cement reinforced coconut fiber is ranged from 8.27 to 13.79 MPa, 9.31 to 14.12 MPa and 10.76 to 15.81 MPa after 7, 14, and 28 days, respectively. The highest MOR of 15.81 MPa was exhibited by the composite with 9 wt. % coconut fiber after 28 days of curing. After 28 days of curing, all composites achieved a highest value of MOR. Although, the reference sample has the highest value of MOR compared to other composite, it is still comparable and the significant difference is small. The change in maximum MOR with change in fiber content could be associated with the fabrication of cement composites. Low fiber content with homogeneous distribution is able to bridge cracks more effectively and introduced balling effect of the fiber. Thus, it leads to better cracks arrest in the composite [17-19]. However, the addition of coconut fiber above 9 wt. % will lead to agglomeration of coconut fiber and reduce the strength of composites.

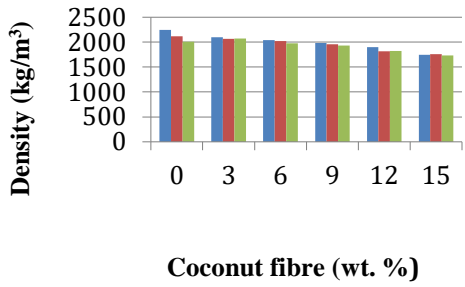

Figure 1a

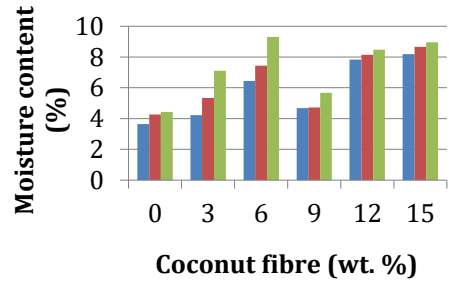

Figure 1b

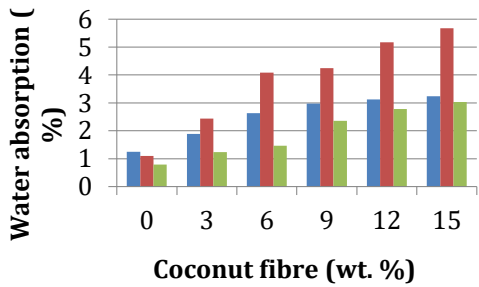

Figure 1c

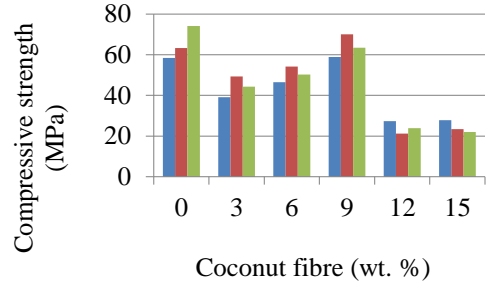

Figure 1d

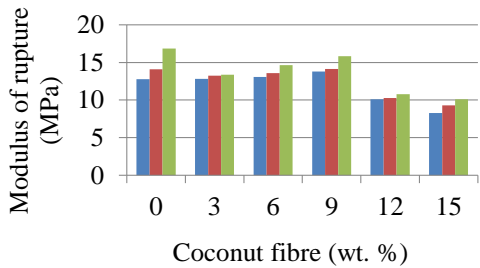

Figure 1e

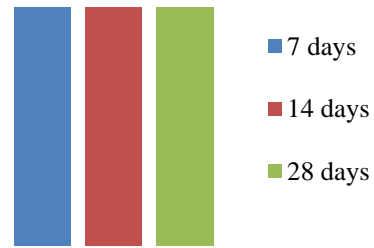

**Figure 1:** Comparison of (a) density (b) moisture content (c) water absorption (d) compressive strength (e) modulus of rupture of cement composite at different percentage of coconut fiber after 7, 14 and 28 days of curing.

## CONCLUSION

In this work, an addition of 15 wt. % of coconut fiber has shown lowest value of density (1735 kg/m$^3$) while reference samples show both lowest values in moisture content (3.64 %) and water absorption (0.79 %) compared to cement reinforced coconut fiber composite. Cement reinforced with 9 wt. % coconut fiber composite gave a better mechanical properties compared to others with highest values in compressive strength (58.98 MPa) and MOR (15.81 MPa). From this research, it was found that the coconut fiber can be used as reinforcement and fly ash can be applied as a substitution of sand in the development of coconut fiber based-green composite. Increasing content of coconut fiber will increase the compressive strength and modulus of rupture until some optimum composition (9 wt. % of coconut fiber).

ACKNOWLEDGEMENTS

The researchers would like to show greatest gratitude to the Ministry of Agriculture and Agro-Based Industry, Malaysia for providing the Research Grant and Universiti Malaysia Perlis for the facilities provided for this research.